\documentclass[12pt]{iopart}
\usepackage{graphicx}
\usepackage{amssymb}
\usepackage{amsthm}
\usepackage{braket}
\usepackage{bbold}
\usepackage{csquotes}
\usepackage{layouts}
\makeatletter
\newcommand{\overset}[2]{\ensuremath{\mathop{\kern\z@\mbox{#2}}\limits^{\mbox{\scriptsize #1}}}}
\makeatother
\def\eqref{\eref}
\def\text{\mathrm}
\newenvironment{align}{\begin{eqnarray}}{\end{eqnarray}}
\newenvironment{multline}{\begin{eqnarray}}{\end{eqnarray}}
\newcommand{\affdn}{\mathcal{A}/(2N)}
\newcommand{\affn}{\mathcal{A}/N}
\newcommand{\aff}{\mathcal{A}}
\newcommand{\imag}{\mathrm{i}}
\newcommand{\bigO}{\mathcal{O}}
\newcommand{\matM}{\mathbf{M}}
\newcommand{\piN}{\frac{2 \pi \imag}{N}}
\newcommand{\piNr}{\frac{2 \pi }{N}}
\newcommand{\freeEn}{\Delta F}
\newcommand{\freeEnT}{\Delta \tilde{F}}

\usepackage{lipsum}
\begin{document}

\title{Affinity-dependent bound on the spectrum of stochastic matrices}% Force line breaks with \\
\author{Matthias Uhl and Udo Seifert}
\address{II. Institut f\"ur Theoretische Physik, Universit\"at Stuttgart, 70550
Stuttgart, Germany}
\date{\today}% It is always \today, today,

\begin{abstract}
	Affinity has proven to be a useful tool for quantifying the non-equilibrium
	character of time continuous Markov processes since it serves as a measure
	for the breaking of time reversal symmetry. It has recently been conjectured
	that the number of coherent oscillations, which is given by the ratio of
	imaginary and real part of the first non-trivial eigenvalue of the
	corresponding master matrix, is constrained by the maximum cycle affinity
	present in the network. In this paper, we conjecture a bound on the whole
	spectrum of these master matrices that constrains all eigenvalues in a
	fashion similar to the well known Perron-Frobenius theorem that is valid
	for any stochastic matrix. As in other studies that are based on
	affinity-dependent bounds, the limiting process that saturates the bound is
	given by the asymmetric random walk. For unicyclic networks, we prove
	that it is not possible to violate the bound by small perturbation of the
	asymmetric random walk and provide numerical evidence for its
	validity in randomly generated networks.
	The results are extended to multicyclic networks, backed up by
	numerical evidence provided by networks with randomly constructed topology
	and transition rates.
\end{abstract}

\section{Introduction}

Real valued matrices with positive entries (with exception of the diagonal) can
be encountered in many different fields of mathematics and physics. They show
up in many forms and under different names throughout the literature. In graph
theory they appear as the Laplacian matrix of (weighted)
graphs~\cite{agaev_spectra_2005}, the time evolution of Markov chains is
governed by a transition matrix that falls under this category, and most
importantly for the scope of this article, the time evolution of a continuous
time jump process is generated by a matrix of this type.  Efforts to understand
the structure of the spectrum of such matrices can ultimately result in
insights into  the studied system.

While many well known results like the Perron-Frobenius theorem or Gershgorin
disks are quite general~\cite{meyer_matrix_2000,maroulas_perron-frobenius_2002, pillai_perron-frobenius_2005},
in a physical context bounds on the spectrum that may be less general but
depend on physically meaningful quantities are more desirable since they could
be used to infer otherwise hidden properties of the system. In particular, for
non-equilibrium systems coupled to thermal or chemical reservoirs such a
strategy is called thermodynamic inference~\cite{seifert_stochastic_2019}.
A recent, prominent example for such a relation is the thermodynamic
uncertainty relation, which provides a lower bound to the rate of entropy
production based on the observable precision of thermodynamic
currents~\cite{barato_thermodynamic_2015,
gingrich_dissipation_2016,
horowitz_proof_2017a,proesmans_discrete-time_2017a}

Recent efforts to understand relations between the entropy production
associated with maintaining biochemical
oscillations~\cite{fei_design_2018,nguyen_phase_2018,owen_number_2019}
have sparked interest in fundamental connections between the non-equilibrium
character of such reactions and the properties of the observed oscillations. It
was conjectured that the affinity of the chemical network can be used to find a
bound to the number of coherent oscillations shown by the dominant contribution
to the corresponding relaxation process~\cite{barato_coherence_2017}.

This finding begs the question, whether the dominant eigenvalue of the
generator that governs the long time behavior is the only one for which
affinity-dependent bounds apply or whether there are global bounds valid for
all eigenvalues, i.e., for
all timescales of the relaxation process. In this paper we argue that indeed
the later is the case and there exist such a bound for the whole spectrum of
the master equation.

This study is concerned with time continuous Markov
processes on a discrete set of $N$ states. The state of the system jumps with rates
$w_{ij} \geq 0$ from state $i$ to state $j$. Consequently, the probability
$p_{i}(t)$ to occupy a certain state $i$ at time $t$ evolves according to the
master equation
\begin{equation}
	\partial_{t} p_{i}(t) = \sum_{j} M_{i,j} p_{j}(t)
\end{equation}
with the generator $M_{i,j}$ that is of the form
\begin{equation}
	M_{i,j} = w_{ji} - \delta_{i,j} r_{i} \,,
\end{equation}
where the exit rate $r_{i}$ is the sum of all rates of jumps away from state
$i$,i.e., $r_{i} = \sum_{j} w_{ij}$.

An important subclass of such networks that will be used as paradigmatic
examples are unicyclic networks where the states are arranged in a cyclical
fashion and only jumps between next neighbors are allowed. The
generator then takes the form
\begin{equation}
	M_{i,j} = w_{i+} \delta_{i,j+1} +w_{i-}\delta_{i,j-1}  -(w_{i+}+w_{i-})
	\delta_{i,j} \,,
	\label{eq:unicyclicMaster}
\end{equation}
where we assume circular boundary conditions in the indices, i.e., we identify
$N+1 \hat{=} 1$.

The objective of this article is to motivate and conjecture a bound that
interpolates between the generic case covered by the Perron-Frobenius theorem
and the special case of thermal equilibrium, where detailed balance holds.

\section{Conjecture}
\label{sec:conjecture}
\begin{figure}[tpb]
\begin{center}
	\includegraphics[scale=1]{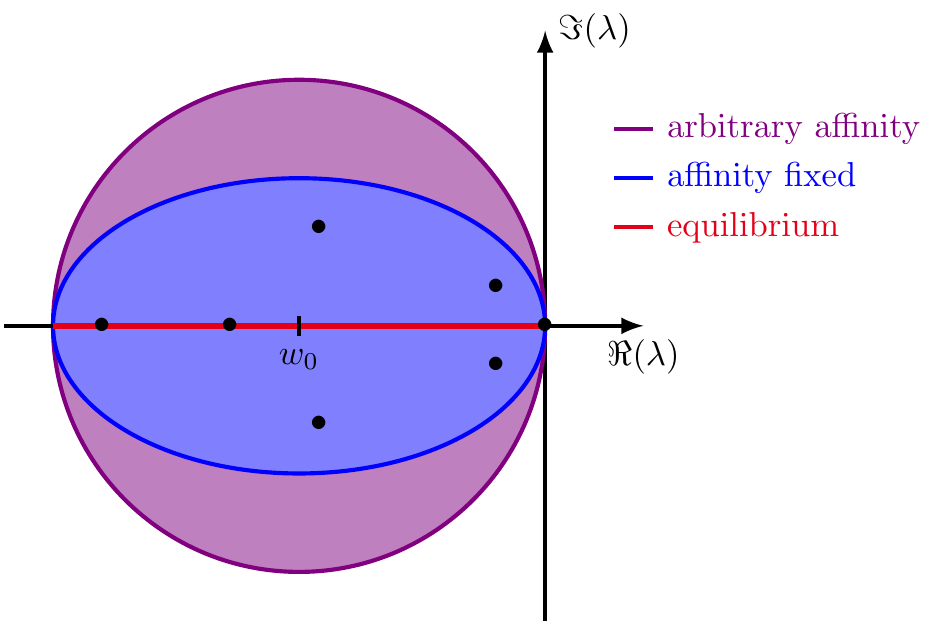}
\end{center}
\caption{Scheme of regions in the complex plane within which
the eigenvalues of a master matrix must lie. For generic master generators the
Perron-Frobenius theorem restricts the eigenvalues to a circle on the negative
half plane. If the system is in thermal equilibrium, in addition, all
eigenvalues must be real. We conjecture an elliptical bound
that depends on the maximum cycle affinity per state.}
\label{fig:bound_schematics}
\end{figure}

An asymmetric random walk on a cycle of states is uniquely defined by the forward
rate $w_{+}$, the backward rate $w_{-}$, and the number of states $N$. It can
alternatively be defined using the exit rate $w_{0}  = w_{+} + w_{-}$, the
affinity $\aff = N \ln(w_{+}/w_{-})$, and $N$.
The corresponding generator reads
\begin{equation}
	\matM_{i,j}^{(0)} = w_{+}\delta_{i,j+1} + w_{-}\delta_{i,j-1} - (w_{+}
	+w_{-})\delta_{i,j}
\end{equation}
with the rates
\begin{equation}
	w_{+} = w_{0} \frac{e^{\aff/(2N)}}{2 \cosh(\aff/(2N))} \quad \mathrm{and} \quad
	w_{-} = w_{0} 	\frac{e^{-\aff/(2N)} }{2 \cosh(\aff/(2N))}\,.
\end{equation}
It is a special case of a unicyclic system with $w_{i+}$ and $w_{i-}$ in
equation~\eqref{eq:unicyclicMaster} chosen uniformly for each link.
This matrix is circulant and as such it can be diagonalized analytically
leading to the eigenvectors
\begin{equation}
	\ket{\nu^{0}_{n}} =\frac{1}{\sqrt{N}}  \sum_{k=0}^{N-1} \exp(-2 \pi \imag n
	k /N) \ket{k}\,,
\end{equation}
with the corresponding eigenvalues
\begin{equation}
	\lambda_{n} =w_{0} \left[  -1 + \cos(2 \pi n/N) + \mathrm{i} \tanh(\aff/2N) \sin(2
\pi n/N) \right]\,.
\end{equation}
They lie on an ellipse on the complex plane. We conjecture that the eigenvalues of
the generators of all unicyclic processes that
have the same affinity, defined as $\aff = \sum_{i} \ln
(w_{i+}/w_{i-})$, the same maximum exit rate $w_{0} = \max_{i}  (w_{i+} +
w_{i-})$, and the same number of states, lie
within the ellipse defined by the corresponding asymmetric random walk, as it
is illustrated in figure~\ref{fig:bound_schematics}.
For multicyclic networks, we conjecture that the eigenvalues lie within
the ellipse corresponding to the cycle $\mathcal{C}$ present in the network
that maximizes the ratio $\aff_{\mathcal{C}}/N_{\mathcal{C}}$, where
$\aff_{\mathcal{C}}$ and $N_{\mathcal{C}}$ denote the affinity and the number
of states contained in cycle $\mathcal{C}$, respectively.

\section{Motivation}%
\label{sec:motivation}

Having stated the conjecture, we provide some rationale as to why this should
be the case.
If the system is in equilibrium, the transition rates have to satisfy detailed
balance relations that connect the transition rates to the
free energy $\mathcal{F}_{i}$  associated with the states according
to~\cite{seifert_stochastic_2019}
\begin{equation}
	\frac{w_{ij} }{w_{ji} }  = \exp \left( \mathcal{F}_{j}  -\mathcal{F}_{i}
		\right)\,.
\end{equation}
A suitable set of free energies can be found if and only if the cycle affinity
\begin{equation}
	\aff_{\mathcal{C}}  = \sum_{i \rightarrow j \in \mathcal{C}} \ln
	\frac{w_{ij}}{w_{ji}}\,
\end{equation}
of all cycles in the network vanishes. As a consequence of the
detailed balance relation, the generator satisfies the symmetry relation
\begin{equation}
	\exp(\mathcal{F}_{j} /2)M_{j,i} \exp(-\mathcal{F}_{i}/2)  = \exp(\mathcal{F}_{i}/2)  M_{i,j} \exp(-\mathcal{F}_{j}/2)
\end{equation}
and is thus similar to a symmetric matrix which means that its spectrum is
real. This result agrees with our conjecture when taking the limit $\aff
\rightarrow 0$ in which case the ellipse degenerates to a line on the real
axis.

For arbitrary affinities the Perron-Frobenius theorem guarantees that no
eigenvalue of the generator lies outside the circle centered at $-w_{0}$ on the complex plane
 that touches the imaginary axis. This bound corresponds to
our conjecture in the limit $\aff/N \rightarrow \infty$.

An early indication that there is a connection between the distance from equilibrium
and the spectrum
can be found in works of Dimitriev and Dynkin on refinements of the
Perron-Frobenius theorem (originally published
in~\cite{dmitriev_characteristic_1945, dmitriev_characteristic_1946} for a
translation c.f.~\cite{swift_location_1972}).
There, it was proven that the master generators capable of saturating the
Perron-Frobenius bound on the eigenvalues, are up to a multiplicative constant
and permutations of states matrices of the form
\begin{equation}
	M_{i,j}^{\infty}  = \delta_{i,j+1} - \delta_{i,j}\,.
\end{equation}
Remarkably, this is the generator of an asymmetric random walk in the
limit $\aff \rightarrow \infty$, which shows that the affinity must
diverge if the bound from the Perron-Frobenius theorem is to be saturated.

Moreover, related studies that are concerned with bounds to certain physical
quantities like the Fano factor of thermodynamic
currents~\cite{gingrich_dissipation_2016,proesmans_discrete-time_2017a,pietzonka_universal_2016} or the number of coherent
oscillations~\cite{barato_coherence_2017} share several important aspects that can serve as guiding
principles to identify a bound on the spectrum that depends on the
non-equilibrium nature of the process.
\begin{enumerate}
\item The maximum cycle affinity per state, i.e.,
	\begin{equation} \max_{\mathcal{C}} \frac{\aff_{\mathcal{C}}
	}{N_{\mathcal{C}} }   \end{equation} is the quantity of choice to
	characterize the distance of the system from thermo\-dynamic equilibrium.
\item The asymmetric random walk with the desired affinity per state is the
	process that is extremal in the sense of the considered bound, i.e., a
	cyclic process with uniform backwards and forward rates leads to saturation
	of the bound. For example, in the case of the affinity-dependent bounds on the
	Fano factor, it turned out that out of all unicyclic processes with the same
	affinity and number of states, the asymmetric random walk is the one with
	the lowest Fano factor. In a similar vein, the number of coherent
	oscillations turned out to be maximal in the case of an asymmetric random
	walk.
\item  It became evident that these bounds, which where initially
formulated for unicyclic networks, can be generalized to multicyclic networks by
identifying the cycle that corresponds to the weakest bound. The bound
generated by this cycle serves as a global bound for the whole system.
\end{enumerate}

Following this line of reasoning, the conjecture stated in
section~\ref{sec:conjecture} is nothing but the application of these principles
to the complete spectrum of the master equation.
To substantiate the conjecture, we show in
section~\ref{sec:vicinity_of_the_asymmetric_random_walk} for the unicyclic
case, that the eigenvalues corresponding to the asymmetric random walk are
locally optimal in the sense that there exists no perturbation to the
corresponding generator that shifts the eigenvalues outside of the
conjectured bound. In section~\ref{sec:numerical_observations} we present
numerical evidence obtained by numerical diagonalization of generators of
unicyclic systems with randomly generated rates at fixed affinity. We also show
results of a numerical optimization procedure designed to find a violation of
the bound, failing to do so.
Results for randomly generated multicyclic networks are presented in
section~\ref{sec:multicyclic_case}.  They conform with an elliptical bound
obtained by the cycle that has the maximum link affinity.

\section{Perturbation theory around the asymmetric random walk}%
\label{sec:vicinity_of_the_asymmetric_random_walk}
The conjecture implies that the asymmetric random walk
is an extremal process in the sense that the eigenvalues of its generator
lie on the conjectured bound for all other processes with the
same affinity and number of states. The goal of this section is to prove that
it is indeed not possible for the eigenvalues to move outside of the ellipse
defined by the random walk if the random walk is perturbed in a fashion that
preserves the affinity and the topological structure of the network through
second order perturbations. We also show that these perturbations can not
vanish, which means that the ellipse corresponding to the eigenvalues of the
asymmetric random walk can be considered a local optimum of the optimization
problem of finding the least eccentric ellipse that contains all eigenvalues of
a unicyclic generator for a given affinity.
Whether it is also a global bound as we conjecture remains to be proven.

We assume a perturbation of the form
\begin{equation}
	\matM = \matM^{(0)}	+ \epsilon \matM^{(1)}\,.
\end{equation}
Throughout this section we normalize $\matM$ such that the maximum exit rate
takes the value 1, i.e., we set $w_{0} = 1$. This is possible without loss of
generality since we assumed that the maximum exit rate is known. Results for
any value of $w_{0} \neq 1$ can be obtained by rescaling in time.
As shown in appendix~\ref{ssub:perturbation_theory_of_normal_matrices}, the
eigenvalues can be approximated by
\begin{equation}
	\lambda_{n} = \lambda_{n}^{0} + \epsilon \braket{\nu_{n}^{0}| \matM^{(1)}  |
		\nu_{n}^{0}} + \epsilon^2 \sum_{m \neq n} \frac{\braket{\nu_{n}^{0}|
			\matM^{(1)}  |
		\nu_{m}^{0}}\braket{\nu_{m}^{0} | \matM^{(1)}  | \nu_{n}^{0}}
}{\lambda_{n}^{0} - \lambda_{m}^{0}    }   + \bigO(\epsilon^3) \,,
\end{equation}
where $\lambda_{n}^{0}$ and $\ket{\nu^{n}}$ are the $n$-th eigenvalue and
eigenvector of the unperturbed asymmetric random walk, respectively.

\subsection{First order perturbation of the asymmetric random walk}%
\label{sub:first_order_perturbation_of_the_asymmetric_random_walk}

The perturbation has to comply with the following rules
\begin{enumerate}
	\item $M_{i,j}^{(1)}  $ is nonzero only if $M_{i,j}^{(0)} $ is also
		nonzero as we only want to perturb existing rates and not introduce new
		connections between states of the network.
	\item The columns of $M^{(1)}$ must sum up zero, since the resulting matrix
		must still be a Markov generator.
	\item Since we normalized the matrix $\matM$ such that all exit rates are less or
		equal to 1, the diagonal entries of $\matM^{(1)}$ must not be negative and
		$\epsilon$ can only  take on positive values.
	\item The entries of the perturbation matrix have to be chosen in such a
		way that the affinity is preserved under the perturbation.
\end{enumerate}

An ansatz that satisfies the first two constraints is given by the choice
\begin{equation}
	M_{i,j}^{(1)}   = -k_{i+} \delta_{i,j+1}  -k_{i-}  \delta_{i,j-1} +
	(k_{i+} + k_{i-}) \delta_{i,j}\,
\end{equation}
while the third constraint corresponds to the condition
\begin{equation}
	k_{i+} + k_{i-} \geq 0	\quad\forall \, i\,.
	\label{eq:kpos}
\end{equation}
Fixed affinity of the perturbed system translates to
\begin{equation}
	\left(\frac{w_{+} }{w_{-} } \right)^{N}  = \exp \left(\mathcal{A}
		\right) \overset{!}{=} \frac{\prod_{i=0}^{N-1}( w_{+} +\epsilon k_{i,+} )
		}{\prod_{i=0}^{N-1} \left( w_{-} +\epsilon k_{i-}   \right) }
		= \frac{w_{+}^{N}+ \epsilon \Gamma_{+} w_{+}^{N-1}         }{
			w_{-}^{N} + \epsilon \Gamma_{-} w_{-}^{N-1}   }  +
			\mathcal{O}(\epsilon^2)
\end{equation}
For perturbation theory of first order this condition needs only to be
satisfied up to first order, which reads
\begin{equation}
	\left(\frac{w_{+}}{w_{-} } \right)^{N}   =	\left(\frac{w_{+}}{w_{-} }
	\right)^{N} + \epsilon \frac{\Gamma_{+} w_{+}^{N-1} w_{-}^{N} -
	\Gamma_{-} w_{-}^{N-1} w_{+}^{N}     }{w_{-}^{2N}  }
	+\mathcal{O}(\epsilon^2)
	\label{eq:stoerungAffBed}
\end{equation}
with the sum over all perturbations of all forward or backward rates defined as
$\Gamma_{\pm} \equiv \sum_{i} k_{i\pm}$. Satisfying  these conditions fixes the
ratio of the two sums to
\begin{equation}
	\frac{\Gamma_{+}}{\Gamma_{-}} = \frac{w_{+} }{w_{-} }\,.
	\label{eq:ratio}
\end{equation}

According to the perturbative solution derived
in~\ref{ssub:perturbation_theory_of_normal_matrices}, the
perturbation to the
eigenvalues are in first order given by
\begin{eqnarray}
\fl	\braket{\nu^{n} | M^{(1)} | \nu^{n} } &= \frac{1}{N} \sum_{m,k} \exp(2
	\pi \mathrm{i} (m \! - \! k) n /N) \left( \delta_{m,k} ( k_{k+} + k_{k-} )
	\! - \! k_{k+} \delta_{m,k+1}\! - \! k_{k-} \delta_{m,k-1}  \right)\nonumber\\
	&= \frac{1}{N} \sum_{k} (k_{k+} + k_{k-}) - k_{k+} \exp(2 \pi
	\mathrm{i} n/N)   -k_{k-} \exp(-2 \pi \mathrm{i} n/N) \nonumber \\
	&= \frac{1}{N} \left[ \Gamma_{+} + \Gamma_{-} - \Gamma_{+} \exp (2
\pi \mathrm{i} n/N) - \Gamma_{-} \exp(-2 \pi \mathrm{i} n /N )     \right] 
\label{eq:firstOrder}
\end{eqnarray}
in which only the sums
$\Gamma_{\pm}$ appear. This, in combination with the fact that the ratio
between the two sums is fixed as per equation~\eqref{eq:ratio}, makes it
possible to treat every perturbation that conforms with the constraints with one
relation by introducing the mapping
\begin{equation}
	\Gamma_{+} = \Gamma \exp(\mathcal{A}/(2N))  \quad \text{and} \quad
	\Gamma_{-} = \Gamma \exp(-\mathcal{A}/(2N)) 
\end{equation}
defining the rate constant $\Gamma$, which is the only parameter relevant for
the first order correction and needs to be non negative to satisfy
eq.~\eqref{eq:kpos}.

By inserting the mapping into eq.~\eqref{eq:firstOrder} and applying
trigonometric relations, one finds that  the perturbation can be put into the
rather simple form
\begin{equation}
\fl	\braket{\nu^{n} | M^{(1)} | \nu^{n} } = \frac{2 \Gamma}{N} \left[ \cosh \left( \frac{\mathcal{A}}{2N}  \right) \left(
		1- \cos \left( \frac{2 \pi n}{N}  \right) \right) - \mathrm{i} \sinh
	\left( \frac{\mathcal{A}}{2N}  \right) \sin \left( \frac{2 \pi n}{N}
\right) \right] \,,
\end{equation}
which is in fact always a multiple of the unperturbed eigenvalue allowing us to
write
\begin{equation}
	\lambda_{n} = \left[1-\epsilon \frac{2 \Gamma}{N} \cosh \left(
	\frac{\aff}{2N}   \right)  \right]\lambda_{n}^{0} + \bigO(\epsilon^2)\,.
	\label{eq:firstOrderRes}
\end{equation}
Note that the prefactor in square brackets cannot exceed $1$ since both
$\epsilon$ and $\Gamma$ are non-negative.
This result has the interesting consequence that the first order perturbation always
shifts the eigenvalue towards the origin of the complex plane. This statement
is generically true, with the possible exception that the first order perturbation
vanishes, i.e., for $\Gamma=0$.
Not only does this confirm that it is not
possible  to leave the ellipse defined by the eigenvalues of the asymmetric
random walk, it also confirms the bound conjectured in
ref.~\cite{barato_coherence_2017} in first order around an asymmetric random
walk. There, it was conjectured that the dominant non-zero eigenvalue is
contained within a cone spanning from the origin to the corresponding
eigenvalue of the asymmetric random walk with the same affinity, i.e.,
\begin{equation}
	-\frac{\Im(\lambda_{1} )}{\Re(\lambda_{1})} \leq -
	\frac{\Im(\lambda_{1}^{0})}{\Re(\lambda_{1}^{0})  }  \,.
\end{equation}
From eq.~\eqref{eq:firstOrderRes} it is obvious that this bound is saturated
for first order perturbations. In contrary to the conjecture
in~\cite{barato_coherence_2017}, this result is not limited to the first
non-trivial eigenvalue but holds for arbitrary $n$.

\subsection{Second order perturbation in case of vanishing first order}%
\label{sub:second_order_correction_in_case_of_vanishing_first_order}
While the first order perturbation always points inside the conjectured bound, it is
not guaranteed that it is nonzero. In this section we want to study cases in
which the first order vanished and the second order becomes the dominant one
for small $\epsilon$.

The first order vanishes if and only if $\Gamma=0$. As the perturbations to the rates
$k_{i}$ must also satisfy condition \eqref{eq:kpos}, this implies that
$k_{i+} + k_{i-} =0 $ must hold individually for all $i$, which means that the
perturbation is not allowed to change the exit rates of the asymmetric random
walk.

Rather than expanding the Taylor series of the affinity in
eq~\eqref{eq:stoerungAffBed} up to second order and to derive a further
condition on the perturbation, we opt to include the fixed affinity condition
directly into a suitable ansatz for the transition rates. The ansatz is given
by the choice
\begin{equation}
\fl	w_{i+} = \frac{\exp(\affdn + \epsilon f_{i} )}{2 \cosh (\affdn + \epsilon
	f_{i}) } \quad \text{and} \quad w_{i-} = \frac{\exp(-\affdn - \epsilon
f_{i}) }{2 \cosh(\affdn + \epsilon f_{i}) } = 1- w_{i+}\,,
\label{eq:secondOrderAnsatz}
\end{equation}
where the parameters $f_{i}$ characterize the perturbation and $\epsilon$ is
a small amplitude. Forward and backward rate sum up to unity
as it is necessary for a vanishing first order perturbation.
In order to keep the affinity fixed at $\aff$, the parameters $f_{i}$ must sum
up to zero, i.e.,
\begin{equation}
	\sum_{i} f_{i} = 0  \,.
\end{equation}

Expanding the master matrix corresponding to our ansatz in a Taylor series up
to second order leads to
\begin{equation}
	\mathbf{M} = \mathbf{M}^{(0)}  + \epsilon \mathbf{M}^{(1)} + \epsilon^2
	\mathbf{M}^{(2)}   + \bigO(\epsilon^3)
\end{equation}
with the  perturbation matrices
\begin{equation}
	\matM_{i,j}^{(1)}  = a f_{j}
	\left( \delta_{i,j+1} - \delta_{i,j-1}   \right)
\end{equation}
and
\begin{equation}
	\matM_{i,j}^{(2)}  = -b f_{j}^{2} \left( \delta_{i,j+1} -
	\delta_{i,j-1}   \right)
\end{equation}
containing the constants
\begin{equation}
	a \equiv \frac{2\exp(\affn)}{\left( \exp(\affn) +1 \right)^2} 
	\quad \mathrm{and} \quad
	b \equiv - \frac{2\exp(\affn) \left( \exp(\affn) - 1 \right)}{
	\left( \exp(\affn) +1  \right)^3} \,.
	\label{eq:secondOrderConstants}
\end{equation}

Up to second order in $\epsilon$ the eigenvalues of the matrix $\matM$ are given by
the expression
\begin{equation}
	\fl \lambda_{n} = \lambda_{n}^{0}  + \epsilon \braket{\nu_{n}^{0}  | \matM^{(1)}
	| \nu_{n}^{0} }   
	+ \epsilon^2 \!\! \left( \sum_{m \neq n}  \frac{\braket{\nu_{n}^{0}  | \matM^{(1)}|
			\nu_{m}^{0} } \! \braket{\nu_{m}^{0}  | \matM^{(1)}  | \nu_{n}^{0}  }
			}{\lambda_{n}^{0}  -
	\lambda_{m}^{0}   }  + \braket{\nu_{n}^{0}  | \matM^{(2)} | \nu_{n}^{0}  }
\!\!\right) +
	\bigO(\epsilon^3).
	\label{eq:secondOrder}
\end{equation}
We proceed by calculating the matrix elements of the perturbation matrices in
the eigen-basis of the unperturbed system. For $\matM^{(1)}$ we find
\begin{eqnarray}
	\braket{\nu_{n}^{0}  | \matM^{(1)}  | \nu_{m}^{0} }   &= \frac{1}{N} \sum_{k,l}
	\exp \left[  \piN (k n - l m ) \right ] \matM_{k,l}^{(1)}   \nonumber\\
	&= \frac{1}{N} \sum_{k,l}   \exp \left[  \piN (k n - l m) \right ]  a
	f_{l} \left( \delta_{k,l+1} - \delta_{k,l-1}   \right)  \nonumber\\
	&= \frac{a}{N} \left(   e^{\piN n} - e^{-\piN n}  \right)
\sum_{l}  \exp \left[ \piN l
(n-m) \right] f_{l}\,.
\end{eqnarray}
Since the eigenvectors only contain terms of the form $e^{\piN n m}$, the
discrete Fourier-transform defined as
\begin{equation}
	\tilde{f}_{i} \equiv  \frac{1}{\sqrt{N}} \sum_{k} e^{- \piN k i} f_{k}\,,
\end{equation}
arises naturally, which allows us to write
\begin{equation}
	\braket{\nu_{n}^{0}  | \matM^{(1)}  | \nu_{m}^{0}  }\braket{\nu_{m}^{0}  |
		\matM^{(1)}  |
	\nu_{n}^{0} } =- \frac{4 a^2}{N} \sin \left(\piNr n\right) \sin \left( \piNr
m \right) \left| \tilde{f}_{n-m}  \right|^2\,,
\label{eq:secondOrderFirst}
\end{equation}
where we have used the symmetry relation $\tilde{f}_{-i} = \tilde{f}_{i}^{*} $ that holds
because all $f_{i}$ are real.

Following analogous calculations, the matrix elements of $\matM^{(2)}$ entering
eq.~\eqref{eq:secondOrder} read
\begin{align}
	\braket{\nu_{n}^{0}  | \matM^{(2)} | \nu_{n}^{0} } &= \frac{1}{N} \sum_{k,l} \exp
	\left[ \piN n \left( l-k \right) \right] M_{k,l}^{(2)}   \nonumber\\
	&= -\frac{b}{N} \left( e^{\piN n} - e^{-\piN n}   \right) \sum_{l}
	f_{l}^{2}\,.
\end{align}

In order to combine the corrections arising form $\matM^{(1)}$ and $\matM^{(2)}$,
we make use of the fact that the discrete Fourier transformation preserves the
norm of the transformed vector, which can be shown as follows
\begin{multline}
	\sum_{l} f_{l}^{2}  = \sum_{l,k} f_{l} f_{k}  \delta_{l,k} = \frac{1}{N}
	\sum_{l,k,m} f_{l} f_{k}  e^{\piN m (k-l)} \nonumber\\
	=\sum_{m} \left(
	\frac{1}{\sqrt{N} }\sum_{l}  f_{l} e^{-\piN lm } \right)  \left(
\frac{1}{\sqrt{N}} \sum_{k} f_{k} e^{\piN km}     \right) = \sum_{m} \left|
\tilde{f}_{m}  \right|^{2} \,.
\label{eq:parceval}
\end{multline}
Furthermore, we know that the $f_{i}$ sum up to zero, so the zeroth Fourier
coefficient must vanish. Combined with eq.~\eqref{eq:parceval} this allows us
to write the correction from $\matM^{(2)}$ in the same form as
eq.~\eqref{eq:secondOrderFirst}
\begin{equation}
	 \braket{\nu_{n}^{0}  | \matM^{(2)} | \nu_{n}^{0} }= - 2 \imag \frac{b}{N} \sin \left( \piNr n \right) \sum_{m \neq
n} \left| \tilde{f}_{n-m}  \right|^2
\label{eq:secondOrderSecond}
\end{equation}

This means that the second order correction can be split up into a sum over
all Fourier components of the perturbation parameters $f_{i}$. The $N$
coefficients $\tilde{f}_{i} $ are, however, not independent. They have to
fulfill the symmetry relation $\tilde{f}_{-i} = \tilde{f}_{i}^{*}$, which, in combination
with the periodic boundary condition $\tilde{f}_{i+N} = \tilde{f}_{i}$, leads to the
result $\tilde{f}_{i} =
\tilde{f}_{N-i}^{*}$. This leaves us with $\lfloor N/2 \rfloor$ independent terms in
the sums in equations~\eqref{eq:secondOrderFirst}
and~\eqref{eq:secondOrderSecond}. The eigenvalues of the perturbed
system can thus be cast in the form
\begin{align}
	\lambda_{n} &= \lambda_{n}^{0} +  \frac{\epsilon^2}{N} \sin \left(
	\piNr n \right) \left( \sum_{m \neq n} \left[- \frac{4 a^2 \sin \left( \piNr m
\right)}{\lambda_{n}^{0} - \lambda_{m}^{0} } - 2 b \imag \right]  \left|
\tilde{f}_{n-m}  \right|^{2} \right)\\
&= \lambda_{n}^{0}  +  \frac{\epsilon^{2}}{N} \sin \left( \piNr n  \right)  \sum_{k=1}^{\lfloor N/2 \rfloor    } z_{n,k} \left|
\tilde{f}_{k}  \right|^{2}
\label{eq:secondOrderFinal}
\end{align}
with the fundamental directions
\begin{equation}
	\fl z_{n,k} = \cases{\frac{4  a^2 \sin \left( \piNr n
 		\right)}{\lambda_{n}^{0} - \lambda_{n-k}^{0}} - 2 b \imag &
		\hspace{-3cm} if $N$ is even and $k= N/2 $   \\ 
		-\frac{4 a^2 \sin \left( \piNr (n-k)
		\right)}{\lambda_{n}^{0} - \lambda_{n-k}^{0} } - \frac{4 a^2 \sin \left( \piNr (n+k)
		\right)}{\lambda_{n}^{0} - \lambda_{n+k}^{0} } - 4 b \imag &
		otherwise.
	 }
\end{equation}

It is important to note that, unlike the first order, the second order cannot
vanish unless all $\tilde{f}_{i}$ are zero, which would mean that there is no
perturbation at all. Consequently, there is no need to consider higher orders
of perturbation to understand the behavior of the eigenvalues in the
vicinity of the asymmetric random walk.

\begin{figure}[tbp]
	\centering
	\includegraphics[scale=1]{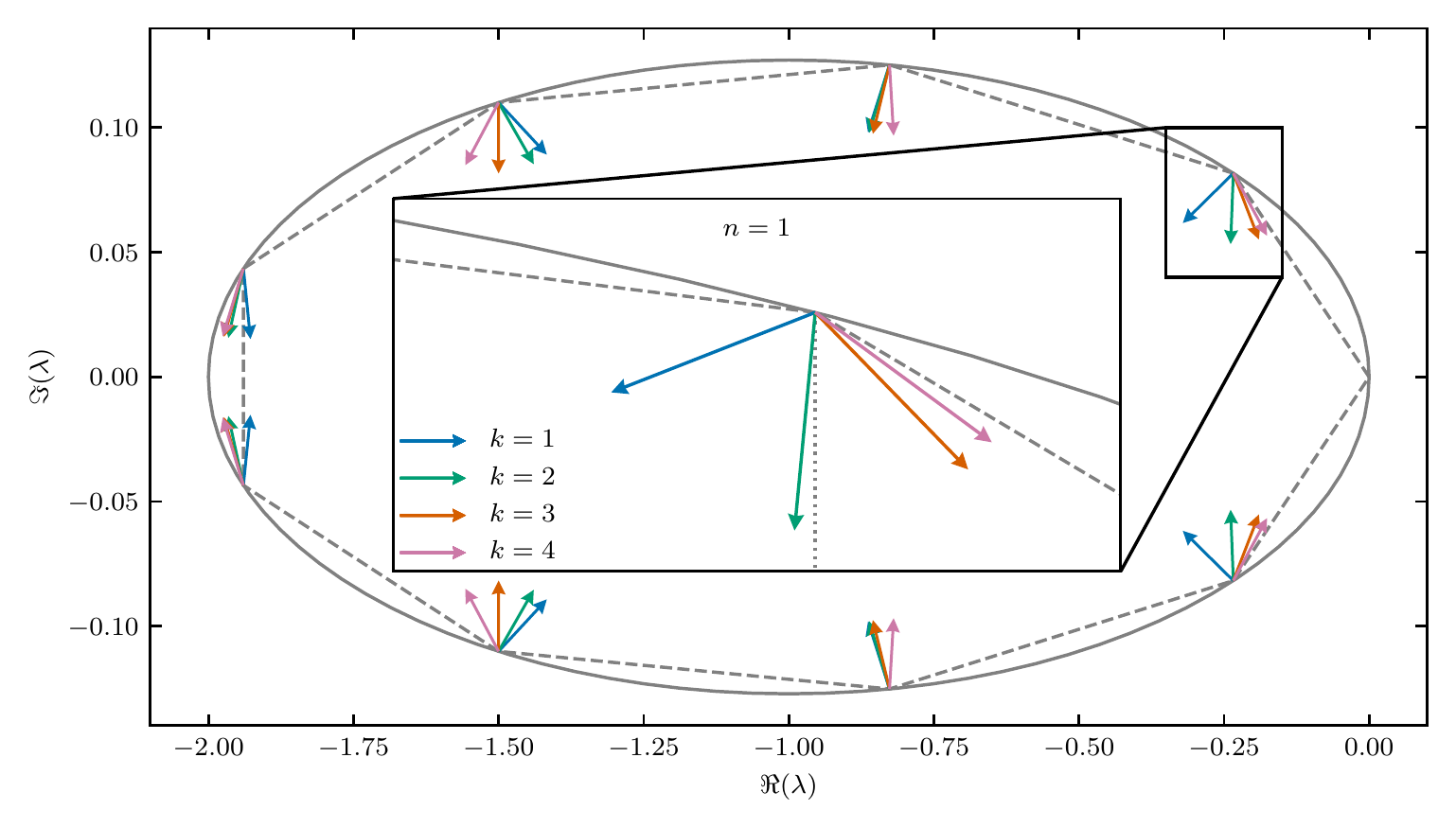}
	\caption{Fundamental directions $z_{n,k}$  of the second order correction to the
	eigenvalues of the asymmetric random walk with $N=9$ states and affinity
$\aff=2.3$. The unperturbed eigenvalues are connected by dashed lines. The
inset shows a magnification  of the first non trivial eigenvalue. All
corrections point to the inside of the ellipse on which the unperturbed
eigenvalues lie. }
\label{fig:secondOrderDirections}
\end{figure}

Figure~\ref{fig:secondOrderDirections} shows the fundamental directions
attached to their respective eigenvalues for a 9 state system. Since the second
order perturbations are a superposition of the fundamental directions, the
eigenvalue can only be shifted in a direction contained in the cone spanned by
two extremal directions. For the eigenvalue highlighted in the inset this cone
is, for example, spanned by the directions $z_{1,1}$ and $z_{1,4}$. In any
case, the entire cone of possible directions points to the inside of the
conjectured bound shown as a gray ellipse.

\begin{figure}[pt]
	\centering
	\includegraphics[scale=1]{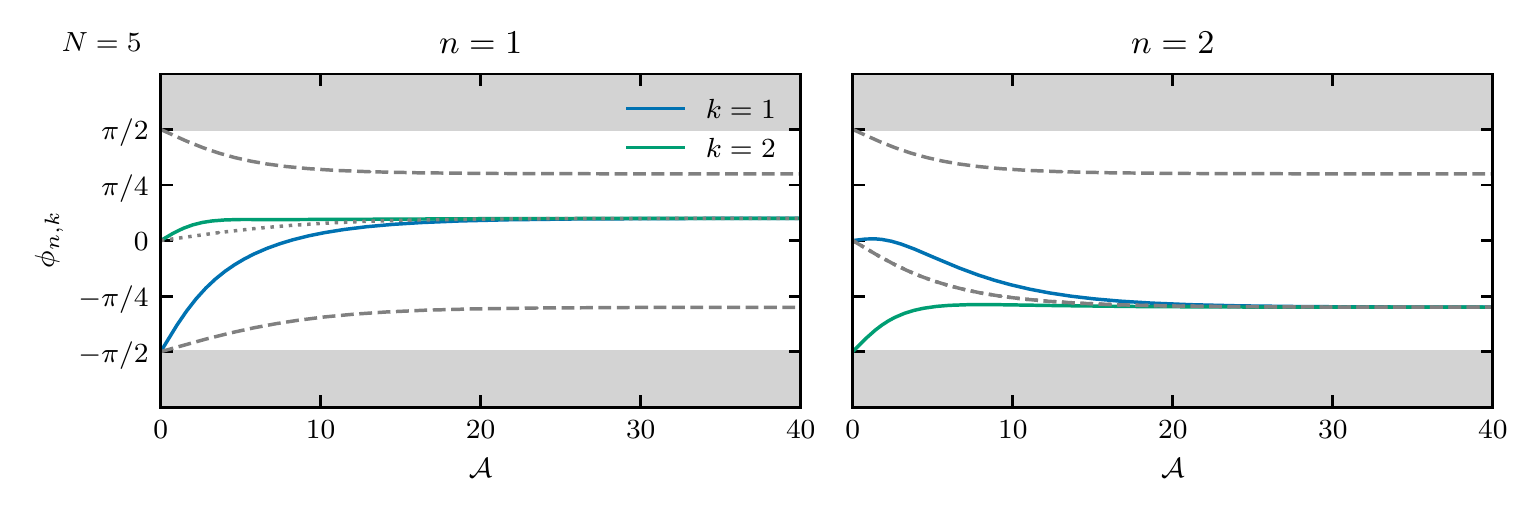}
	\includegraphics[scale=1]{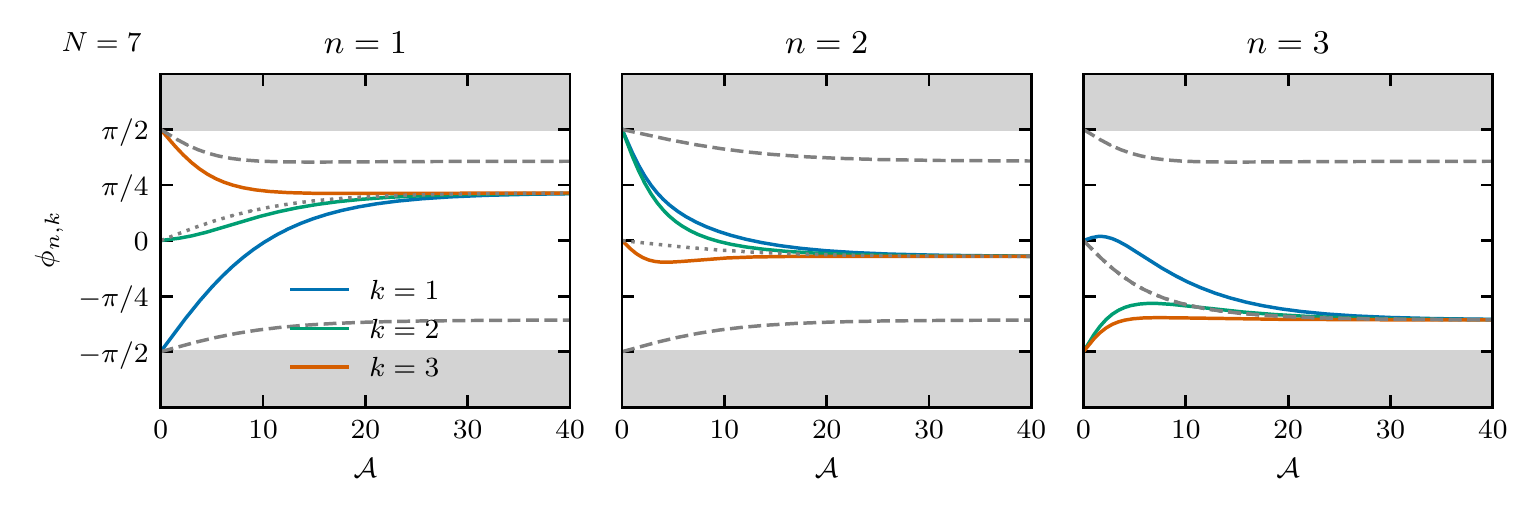}
	\includegraphics[scale=1]{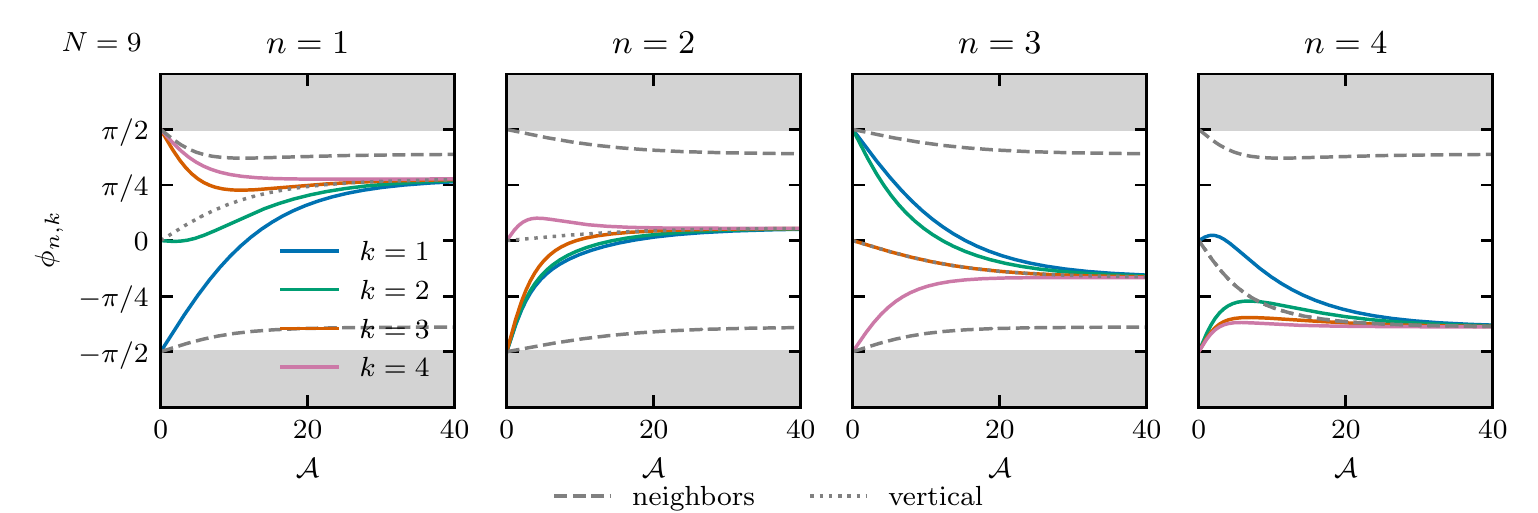}
	\caption{Angles between the fundamental directions of the second order
	perturbation and the direction perpendicular to the ellipse on which the
	unperturbed eigenvalues lie. All possible perturbations are a superposition
of these directions, so the eigenvalue can only move in directions between the
extremal angles. All fundamental direction form an angle between $-\pi/2$ and
$\pi/2$ to the perpendicular direction, showing that there exists no
perturbation that moves the eigenvalues outside of the ellipse. For better
orientation the directions of the neighboring eigenvalues are indicated as
dashed lines. The vertical direction is shown as a dotted line. As the affinity
rises all perturbation directions converge to the vertical direction.}
	\label{fig:angles}
\end{figure}

To better visualize $z_{n,k}$ in the general case, we calculate the
angle on the complex plane between $z_{n,k}$ and
\begin{equation}
	\lambda_{\perp,n}  \equiv -\tanh(\aff/(2N))\cos \left( \piNr n \right)
	-\imag \sin \left( \piNr n \right)\,,
\end{equation}
a complex number that is perpendicular on the ellipse and points to the inside at the
position of the unperturbed eigenvalues. This angle can be
written in the form
\begin{equation}
	\phi_{n,k} = \arcsin \left(\frac{ \Im( z_{n,k} \lambda_{\perp,n}^{*})
	}{|z_{n,k}| |\lambda_{\perp,n}| }   \right)
\end{equation}
and depends only on the affinity and the number of states $N$.
Graphical representations of the affinity-dependence of this angle are
presented for different $N$ in figure~\ref{fig:angles}. For better orientation
the direction of the neighboring eigenvalues and the angle to the downwards
direction are also plotted. It is evident that the angle is always between
$-\pi/2$ and $\pi/2$ meaning that $z_{n,k}$ points inwards. For small
affinities, i.e., close to equilibrium, the fundamental perturbation directions
either point in positive or negative direction or along the imaginary axis, so
the angles are either $\pi/2$, $-\pi/2$ or $0$. Note,
however, that the non degenerate perturbation theory as it is used here is only
valid for $\aff \neq 0$, since all non-trivial eigenvalues of the master matrix
for a random walk become degenerate as $\aff$ approaches zero.

For large affinities $z_{n,k}$ points along the imaginary axis, since in this
case $b \gg a^2$ holds in eq.~\eqref{eq:secondOrderConstants}, while all other terms in eq.~\eqref{eq:secondOrderFinal}
stay finite. For this reason all curves in fig.~\ref{fig:angles} collapse onto
the curve indicating the vertical direction as $\aff$ goes to infinity.

\section{Numerical evidence}%
\label{sec:numerical_observations}

The asymmetric random walk has proven to be the limiting case in established
bounds that depend on the affinity. In order to further substantiate the bound
beyond the second order perturbation theory, in this
section, we will present abundant numerical evidence for this bound in the
unicyclic case by randomly generating rates that lead to a desired value of the
affinity as well as numerical optimization schemes that put the conjectured
bound to the test.

\subsection{Randomly generated unicyclic systems}%
\label{sub:randomly_gnerated_rates}
The rates for the randomly generated unicyclic networks are generated using the ansatz
\begin{equation}
	w_{i+} = w_{i} \exp(\freeEn_{i} /2) \quad \mathrm{and} \quad
	w_{i-} = w_{i} \exp(-\freeEn_{i} /2) \,,
\end{equation}
introducing the timescales $w_{i}$ and the free energy differences between
connected states $\freeEn$. In order to achieve the desired affinity the free
energy differences have  to sum up to $\aff$. For this reason, we first draw
$N$ values $\freeEnT$ independently from a uniform distribution on the interval
$(-0.5,1.5)$ and calculate the energy differences as $\freeEn_{i}  = \aff
\freeEnT_{i}/\sum_{i} \freeEnT_{i}$. The interval is asymmetrical in order to
make divisions by values close to zero less likely, thereby increasing
numerical stability.

The timescales $w_{i}$ are drawn from a uniform distribution between $0$ and
$1$. After calculating all rates, the timescales are adjusted such that the
largest exit rate takes the value $1$. For this reason the width of the
interval from which the $w_{i}$ are drawn is irrelevant.

The results obtained by perturbing the asymmetric random walk show that there
is a fundamental difference between systems with arbitrary exit rates (generic
case) and systems that have uniform exit rates in all states. We therefore also
specifically generate systems with the same exit rate in each state by
rescaling the rates such that this is the case after drawing them as described
above.

For each combination of $N$ and $\aff$ that we study, we draw 10000 sets of
rates both for the generic and for the case of uniform exit rates
and calculate the eigenvalues of the corresponding generator. As an example
the results for $N=5$ and $\aff=1.2$ are presented in
figure~\ref{fig:norm_eigs_scatter_n_7}. The color indicates whether the states
are drawn with random exit rates (blue) or with the same exit rate in each
state (gray). Also shown are the directions of the first order perturbation 
according to
section~\ref{sub:first_order_perturbation_of_the_asymmetric_random_walk}
and the different fundamental directions of perturbations of
second order in case the first order vanishes (c.f.
section~\ref{sub:second_order_correction_in_case_of_vanishing_first_order}).

All eigenvalues lie within the
ellipse defined by the eigenvalues of the asymmetric random walk. As already
conjectured in~\cite{barato_coherence_2017},
the nontrivial eigenvalue with the largest real part always lies below a line
connecting the origin with the first non-trivial eigenvalue of the
corresponding asymmetric random walk. From the numerical data it is evident
that such a relation also holds for the other eigenvalues when compared to their
corresponding eigenvalue of the random walk.
\begin{figure}[tp]
	\centering
	\includegraphics[width=1\linewidth]{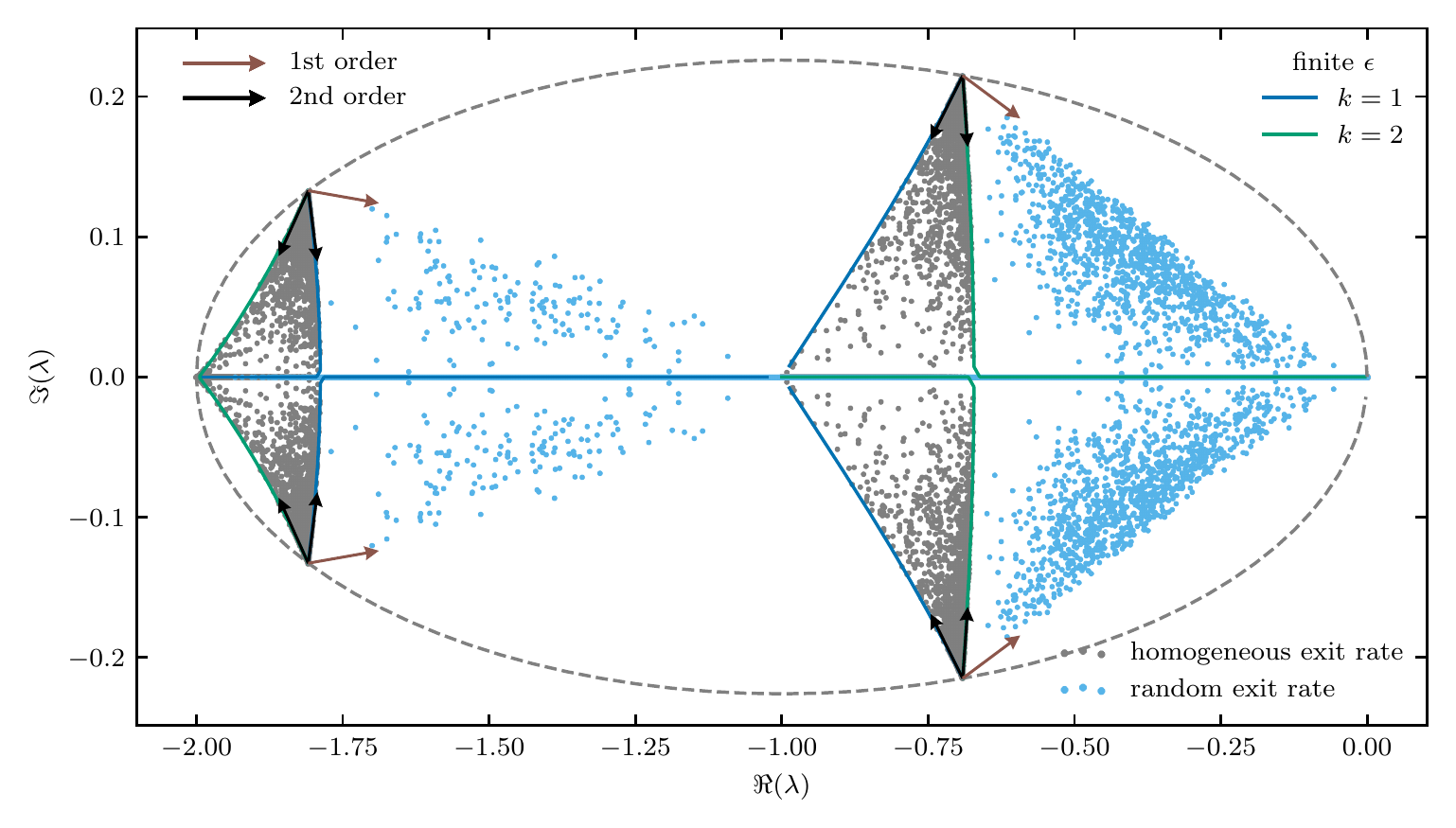}
	\caption{Scatter plot of eigenvalues for systems with $N=5$ states and
		$\aff=1.2$. Random
	systems with unconstrained exit rates are plotted in blue. Systems with
exit rates constrained to 1 are shown in gray. The solid lines show the eigenvalues
of systems with rates constructed with $f_{i}^{(k)}  = \cos(2 \pi i k /N)$ and
finite $\epsilon$.
The arrows indicate the directions of the perturbation results of first and
second order, respectively. From the distribution of eigenvalues it can be seen
that the results obtained by perturbing the asymmetric random walk seem to
extend beyond the validity of the perturbation theory in the sense that the curves
corresponding to the extremal directions of the perturbation theory form
approximate bounds to the eigenvalues not only in the immediate vicinity of the
asymmetric random walk but for all systems with eigenvalues that have an
imaginary part different from zero.}
\label{fig:secondOrderScatter}
\end{figure}

Furthermore, it is interesting to note that the results obtained through
perturbation theory describe the behavior of the eigenvalues rather well, even
if the system is not close to the asymmetric random walk. The eigenvalues of
systems with uniform exit rate stay roughly within the cone derived from the
extremal directions introduced in
section~\ref{sub:second_order_correction_in_case_of_vanishing_first_order}.
This begs the question to which extend the second order perturbation derived
there constitutes a good approximation for large $\epsilon$.

To investigate this
further, we calculate the eigenvalues of unicyclic systems with rates as
defined in equation~\eqref{eq:secondOrderAnsatz} for finite $\epsilon$ numerically.
As we have seen, perturbations that correspond to the fundamental directions of
the second order result are produced by the choice of $f_{i}$ such that its
discrete Fourier-transform only contains one selected mode. For this reason we
choose $f_{i} ^{(k)} = \cos(2 \pi i k /N)$.
The results are shown in figure~\ref{fig:secondOrderScatter} as solid lines
parametrized by $\epsilon \in [0,10]$. By definition these curves leave the eigenvalues
in parallel to the fundamental direction corresponding to the selected
Fourier mode. The curves stay approximately linear,  which shows that, in this
case, the perturbative result can qualitatively describe the spectrum of all
systems that show oscillations and therefore have corresponding eigenvalues with
nonzero imaginary part. Even for larger $\epsilon$, where the non-linearity
becomes more pronounced, we see that the curves still approximately envelope
the eigenvalues for randomly generated systems.

The process has been repeated for values of $N$ between $3$ and $13$ and values
of $\aff$ in the range between $0$ and $20$ with the same qualitative result
(data not shown).

\subsection{Numerical optimization procedure}%
\label{sub:numerical_optimization_procedure}

While studying randomly generated systems can capture the generic behavior of
the eigenvalues, in this section we describe a numerical method that aims to
find possible violations of the conjectured bound (and fails to do so).
The goal is to use standard numerical optimization algorithms to find a set of
rates that produces the maximum imaginary part of a specific eigenvalue while
keeping its real part fixed at a predetermined value and also obeying the
constraints on affinity and exit rates.

\begin{figure}[tp]
	\centering
	\includegraphics[scale=1]{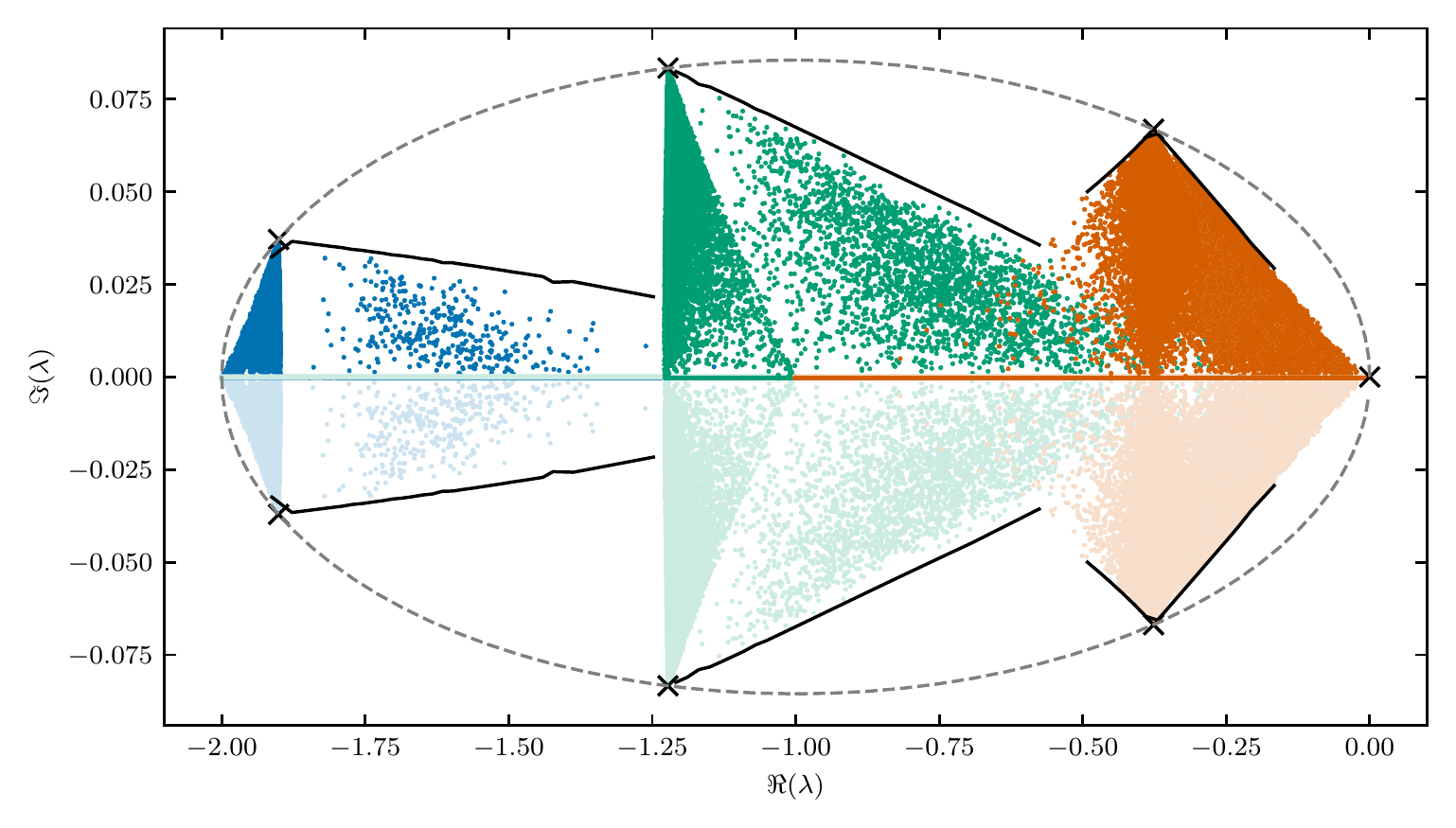}
	\caption{Scatter plot showing the eigenvalues of a cyclic Markov process
		with 7 states and an affinity of $\aff=1.2$.
	The eigenvalues are sorted by increasing real part and are colored accordingly.
	Black curves show results of a numerical optimization scheme with the goal to find
	the process which leads to the largest (smallest) imaginary part of a specific eigenvalue with
the real part fixed at a given value. Black crosses indicate the eigenvalues of
the asymmetric random walk with affinity $\aff$. The numerical results suggest
that it is not possible to find a process at fixed affinity with eigenvalues
that lie outside a certain region that is bounded by the ellipse on which the
eigenvalues of the asymmetric random walk lie.}
	\label{fig:norm_eigs_scatter_n_7}
\end{figure}

This is done using the sequential least squares programming algorithm as it is
implemented in the python library for scientific
computing scipy~\cite{kraft_software_1988a}, with a
randomly chosen initial guess for the rates. As it is to be expected for a
heavily constrained non-linear optimization problem, the algorithm converges
rather poorly without case specific tweaking of the optimization parameters and
the initial guess. Nevertheless it is possible to find extremal cases in some
intervals of the real part, typically in the vicinity of the eigenvalues of
the asymmetric random walk. The results are depicted as solid lines in
figure~\ref{fig:norm_eigs_scatter_n_7}. They show that the eigenvalues can not
leave the ellipse defined by the random walk even if the rates are
specifically designed to do so. Apparently it is not even possible to exceed
the straight line connecting the origin to the corresponding eigenvalue of the
asymmetric random walk, which shows once more that the results conjectured
in~\cite{barato_coherence_2017} can be extended to subdominant eigenvalues.

\section{Multicyclic case}%
\label{sec:multicyclic_case}

In previous studies of affinity-dependent
bounds~\cite{pietzonka_universal_2016,barato_coherence_2017}, it turned out that the results
obtained for unicyclic systems can be extended to multicyclic systems since
the coupling of two or more cycles to each other can be handled by
identifying the cycle that produces the weakest bound. In our case, the weakest
bound is based on the cycle that maximizes the ratio
$\aff_{\mathcal{C}}/N_{\mathcal{C}} $ of cycle
affinity and number of states in the cycle. In this section we provide
numerical evidence that the elliptical bound can be extended to multicyclic
networks following this rationale.

\subsection{Simple multi cyclic network}%
\label{sub:simple_multi_cyclic_network}

\begin{figure}[tpb]
	\centering
	\includegraphics[scale=1]{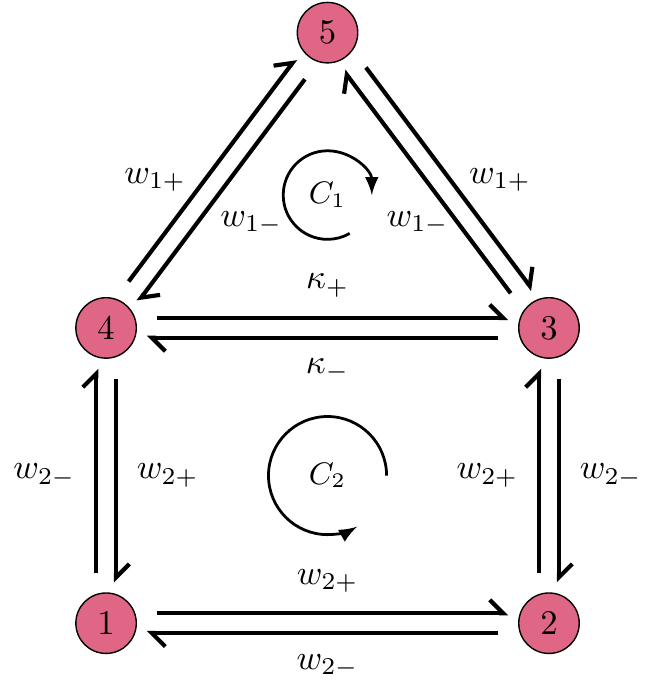}
\caption{Schematic depiction of house network consisting of two fundamental
	cycles. There is a total of three possible cycles. Each
cycle on its own implies an elliptical bound the eigenvalues depending on its
cycle affinity. The weakest of these bounds is generated by the cycle with the
maximum value of $\aff_{\mathcal{C}}/N_{\mathcal{C}}$ and holds even the cycles
are coupled.}
\label{fig:house_tikz}
\end{figure}

 One of the arguably
simplest models to study multicyclic behavior is a network consisting of merely
two fundamental cycles that share one link. Figure~\ref{fig:house_tikz} shows
an example of such a network that was previously used to illustrate universal
bounds on the cumulant generating function of the distribution of entropy
production (\enquote{house network})~\cite{pietzonka_universal_2016}. In this
section we aim to study the influence of coupled cycles on the bounds discussed
above.

As this example is intended to serve as a case study, we want to keep the
number of different rates as low as possible, while still maintaining the
possibility to fix the affinity of each cycle individually and having a
meaningful parameter $\Omega$ that allows us to select which of the cycles has
the dominant influence on the eigenvalues. For this reason we assume that each
transition that belongs uniquely to one of the two fundamental cycles has the
same forwards and backwards rates, respectively. This leaves us with 3 tuples
of transition rates, $w_{1\pm}$ for transitions of cycle 1, $w_{2\pm}$ for
transitions within cycle 2, and $\kappa_{\pm}$ for the transition coupling the
two cycles.

We want to fix the affinities of the cycles. A suitable ansatz is
\begin{eqnarray}
	w_{1+} &= \frac{1}{w_{0} } \exp[ (\aff_{1} - \aff_{\text{link}}) /4 ]  \quad
		   w_{1-} = \frac{1}{w_{0} } \exp[ -(\aff_{1} - \aff_{\text{link}}) /4 ] \\
	w_{2+} &= \frac{1}{w_{0} } \Omega \exp[ (\aff_{2} +\aff_{\text{link}}) /6 ]   \quad
		   w_{2-} = \frac{1}{w_{0} } \Omega \exp[ -(\aff_{2} +
	\aff_{\text{link}}) /6 ] \\
	\kappa_{+} &= \frac{\max \left\{ 1, \Omega \right\}}{w_{0} } \exp(\aff_{\text{link}} /2) \quad
			   \kappa_{-} = \frac{\max \left\{ 1, \Omega \right\}}{w_{0} } \exp(-\aff_{\text{link}}/2)  \\
\end{eqnarray}
that fixes the affinity of cycle 1 to $\aff_{1}$ and the affinity of cycle
 2 to   $\aff_{2}$. It also introduces the parameter $\Omega$ that determines
 the ratio between the timescales of the two cycles. The rate $w_{0}$ is chosen
 such that the maximum exit rate is 1. This means that if $\Omega=0$ only cycle
 1 is active, while $\Omega \rightarrow \infty$ corresponds to the case in
 which only cycle 2 is active.

\begin{figure}[tpb]
	\centering
	\includegraphics[width=0.48\linewidth]{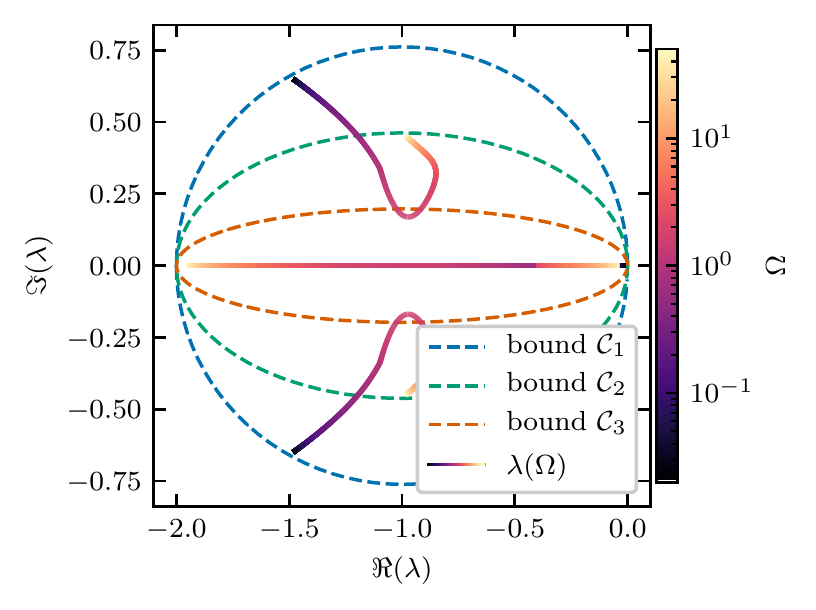}
	\includegraphics[width=0.48\linewidth]{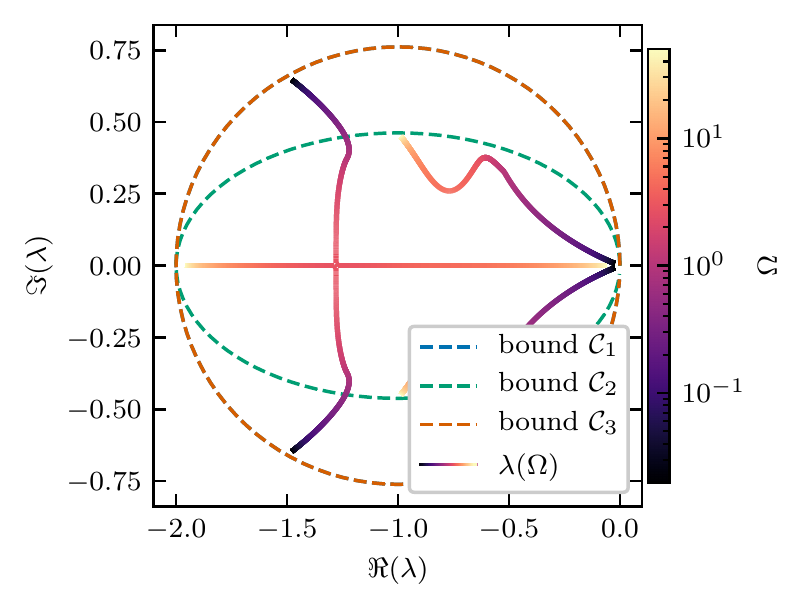}
	\caption{Eigenvalues for the house network as function of $\Omega$ for
		$\aff_{1}=6$ and $\aff_{2}= 4$ (left) and $\aff_{2}=-4$ (right) . The
		elliptical bounds resulting from the average link affinity in each
		cycle are shown as dashed lines. The parameter $\Omega$ is defined such
		that a value of 0 corresponds to an asymmetric random walk with
		affinity $\aff_{1}$ in the upper circle and $\Omega \rightarrow \infty$
		corresponds to an asymmetric random walk with an affinity $\aff_{2}$ in
		the lower circle. For intermediate values the two cycles are coupled.
Nonetheless the eigenvalues stay within the loosest of the three bounds.}
	\label{fig:houseNetwork}
\end{figure}

 Since we already established that the asymmetric random walk is the limiting
 case for unicyclic networks, it is favourable to have this case within our
 parameter space. This can be achieved by making the affinity of the link
 between the two cycles dependent on $\Omega$ in  such a way that it
 interpolates between the values present in a random walk with 3 states and
 affinity $\aff_{1}$ and a random walk with 4 states and affinity $\aff_{2}$. A
 suitable choice is
 \begin{equation}
	\aff_{\text{link}} = \frac{\aff_{1}/3 + \Omega \aff_{2}/4  }{1+\Omega} \,.
 \end{equation}
 Figure~\ref{fig:houseNetwork} shows the resulting eigenvalues as a function of
 $\Omega$ for the affinities $\aff_{1}=6$ and $\aff_{2}=\pm 4$ together with
 the bounds resulting from the average link affinities present in each of the
 three cycles. From the curves it becomes clear that coupling the two cycles
 ($\Omega \simeq 1$) does not drive the eigenvalues outside the loosest
 elliptical bound.
 To a certain extent the behavior of the eigenvalues can be understood
 intuitively. When both affinities are positive (left panel), the two cycles
 sustain oscillations even when coupled, since they do not compete with each
 other on their shared link. The two cycles behave analogously to two
 interlocked gears rotating in opposite directions. For this reason the
 imaginary part of the complex eigenvalue does not vanish for any $\Omega$.
 However, the situation changes when the affinities have different signs (right
 panel). Now, the two cycles compete over the shared link and consequently
 oscillations disappear for intermediary values of $\Omega$ manifesting in real
 valued eigenvalues.

\subsection{Numerical case study}%
\label{sub:numerical_case_study}

While the simple example used in section~\ref{sub:simple_multi_cyclic_network}
can serve as an illustrative case study for a multi-cyclic network, it is by
no means sufficient to make a claim for the generalizability of the unicyclic
bound to arbitrary Markov processes. In this section we will widen the scope of
our case studies to randomly generated networks with fixed number of states but
randomly generated network topology.

Because the loosest bound is produced by the cycle with the largest affinity
per link and the topology is random, it is rather challenging to prescribe the
desired bound and then generate networks randomly that should satisfy this
bound as we did in the previous case studies. Instead, we opt for the reversed
approach, drawing the network first without constraints and computing the
corresponding bound afterwards.

A network is generated by first drawing for each state the numbers of
connections it should have from the uniform distribution of  integers in the
range $[2,N-1]$. This procedure guarantees that  there are no dead ends and
each state is part of at least one cycle. If the configuration is feasible, a
graph with this configuration is generated deterministically and
randomized afterwards by swapping target and source states of randomly selected
connections. The specific procedure is implemented using the graph-tool python
package~\cite{peixoto_graph-tool_2017}. The rates for each link are
drawn from the exponential distribution
\begin{equation}
	p(w_{i}) = \exp(-w_{i})
\end{equation}
Note that the overall timescale, which could be fixed using a factor in the
exponent, is irrelevant in this case, since the rates are normalized afterwards
such that the largest exit rate takes the value $1$.

Networks generated in this manner do, in general, not share the same elliptical
bound, since the cycle affinities and even the cycles themselves are different
for each network. To check whether the bound is satisfied, we iterate over all
cycles contained in the network using the algorithm described
in~\cite{CSTN-013}  and
identify the maximum of $\aff_{\mathcal{C}}/N_{\mathcal{C}}$.

\begin{figure}[tp]
	\centering
	\includegraphics[scale=1]{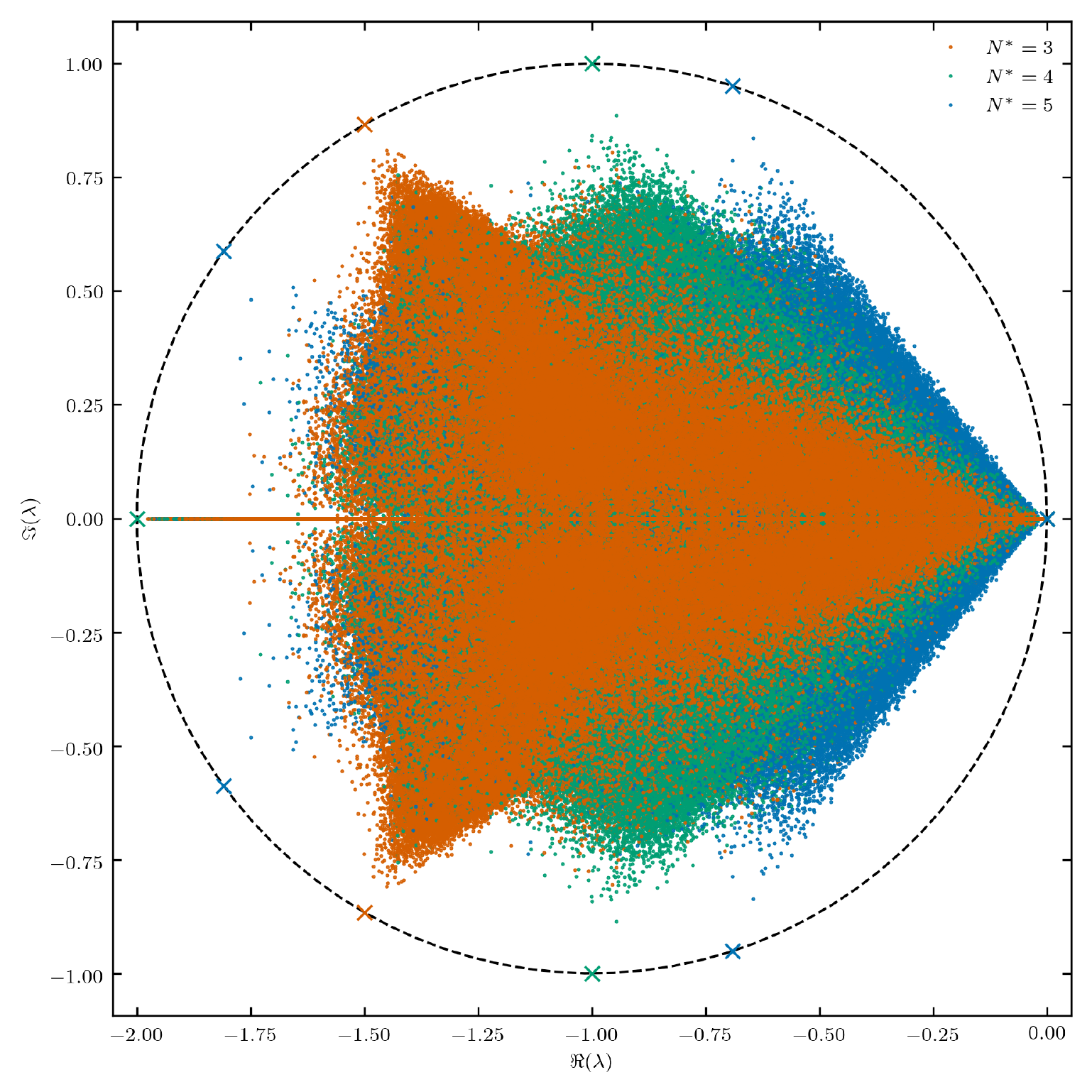}
	\caption{Scatter plot of eigenvalues of randomly generated networks with
	$N=5$ states. The colors encode the length $N^{*}$ of the cycle with
	maximum link affinity.  The imaginary parts of all eigenvalues are rescaled by the
factor of  $(\tanh(\aff^{*} /(2N^{*} ))^{-1}$ in order to make the elliptical
bounds of different networks comparable since these bounds collapse onto the unit
cycle. All eigenvalues lie within the unit cycle, which allows us to conjecture
that the elliptical bound is also valid in the multi cyclic case. }
	\label{fig:scatter_multi_N_5}
\end{figure}

We performed this procedure  $10^{6}$  times for each $N$ up to 7
and did not come across a single violation of the bound. To visualize the
results in a single scatter plot for a given number of states, we scale the
imaginary part of the eigenvalues by a factor of $\left(\tanh \left(
\aff^{*}/N^{*} \right) \right)^{-1} $, which maps all individual bounds
valid for each system to the unit circle around $-1$, thus making networks
with different critical cycles comparable. Figure~\ref{fig:scatter_multi_N_5}
shows the result for $N=5$. The different colors indicate the length  $N^{*}$
of the cycle responsible for the bound.

\section{Application to time discrete Markov chains}%
\label{sub:time_discrete_markov_chains}
While the term affinity is commonly used only for time continuous Markov
processes, our results are not limited to this specific case. Just as the
Perron-Frobenius theorem, they are equally applicable to time discrete Markov
jump processes that are described by the evolution of the probability
$p(i,n)$ to be in state $i$ at the discrete time step $n$ according to
\begin{equation}
	p(i,n+1) = \sum_{j} p(i|j) p(j,n)
\end{equation}
with the transition probability $p(i|j)$.

The definition of the affinity of a cycle translates to the logarithmic
ratio of the probability to observe a forward trajectory along this cycle and
the probability to observe the time-reversed one, i.e.,
\begin{equation}
	\aff_{\text{disc}}  =  \sum_{i \rightarrow j}  \ln \left( \frac{p(j|i)}{p(i|j)}  \right)\,,
\end{equation}
where the sum runs over all links from $i$ to $j$ of the specific cycle.

The eigenvalues of the propagator of a unicylic system with equal forward and
backward jump probability, respectively that is the counterpart of a  time
continuous random walk, are given by
\begin{equation}
	\lambda_{\text{disc},n}  =  \lambda_{n}^{(0)}/w_{0}  +1=   \cos(2 \pi n/N) + \mathrm{i} \tanh(\aff/2N) \sin(2
\pi n/N) \,.
\end{equation}

In regards to the elliptical bound, the treatment of time discrete processes
has the advantage that the timescale, which is explicitly present as $w_{0}$ in
the time continuous case, is inherently fixed by the discreteness of the
process. As a result, the bound does no longer rely on the knowledge of a
parameter other than the affinity per state, which could prove useful for
inferring the discrete affinity from measurements of correlation functions.

In a similar way, the results obtained here could be generalized to arbitrary
mathematical spectral problems of non-negative matrices, although the quantity
that serves as the effective affinity in these cases may not have a physical
interpretation as it is the case for Markovian systems.

\section{Conclusion}%
\label{sec:conclusion}

We studied the intricate connection between the spectrum of generators of time
continuous Markovian dynamics and the affinity, a measure for breaking time
reversal symmetry. Based on extensive numerical case studies and results obtained from
perturbation theory, we conjecture that the eigenvalues of such generators can
be constrained by the ellipse on which the eigenvalues of a corresponding
asymmetric random walk lie.

While we could not provide a rigorous proof for the conjecture, our results
obtained for perturbations of the asymmetric random walk became a formal proof
if it was possible to show that a unicyclic process with homogeneous
rates is the only process with eigenvalues on the conjectured elliptical bound.

The presented results show once again that for many aspects that involve
non-equilibrium systems, affinity-dependent bounds can be obtained by comparing
the system with the corresponding asymmetric random walk with the same affinity
per state. This strategy has already proven useful in deriving
affinity-dependent bounds on distributions of stochastic currents of which the
thermodynamic uncertainty relation is the most prominent example.

Besides a rigorous proof of the conjecture put forth in this work, the most
pressing
open question is whether there is an underlying connection between these
results that would explain why the asymmetric random walk appears time and time
again as the limiting case when quantifying the influence of the distance from
equilibrium to certain physical quantities.

\appendix
\section{Perturbation theory of normal matrices}%
\label{ssub:perturbation_theory_of_normal_matrices}

Since generic master generators are not Hermitian and their eigenvectors are
therefore in general not pairwise orthogonal, the formulas of time
independent perturbation theory, as they are known from quantum mechanics, are
not directly applicable to these matrices. In the case of an asymmetric random
walk, however, the corresponding matrix is normal, i. e.,
\begin{equation}
	\left[\matM^{(0)} ,{\matM^{(0)}}^{\dagger}\right] =   0\,.
\end{equation}
As a consequence, the
eigenvectors of $\matM^{(0)} $ form an orthonormal basis of the
cor\-re\-spond\-ing Hilbert space. The
aim of this section is to re-derive known results from quantum mechanics using
only the normal property of the perturbed matrix.

By virtue of being normal, the generator $\matM^{(0)}$ of an asymmetric random
walk satisfies the relations
\begin{eqnarray}
	\braket{\nu_{n}^{0} | \nu_{n'}^{0} } &= \delta_{n,n'}\\
	\matM^{(0)}   \ket{\nu_{n}^{0} }  &= \lambda_{n} \ket{\nu_{n}^{0}  }  \\
{\matM^{(0)}}^{\dagger} \ket{\nu_{n}^{0}} &= \lambda_{n}^{*}  \ket{\nu_{n}^{0} } \,.
\end{eqnarray}

We want to calculate the eigenvalues $\lambda_{n}$ and the normalized
eigenvectors $\ket{\nu^{n}}$ of the perturbed matrix
\begin{equation}
	\matM = \matM^{(0)}  + \epsilon \matM^{(1)} 
\end{equation}
up to second order in $\epsilon$, i.e., we calculate a perturbative solution to
the eigenvalue equation
\begin{equation}
	(\matM - \lambda_{n}) \ket{\nu_{n}} = 0  	
\end{equation}

Since the eigenvectors of the unperturbed problem form an orthornormal basis, it
is possible to express the perturbed eigenvector as a superposition of these,
such that we can write
\begin{equation}
	\ket{\nu_{n}} = c_{n} \ket{\nu_{n}^{0}}   + \sum_{m \neq n} d_{m}
	\ket{\nu_{m}^{0}}      \,,
\end{equation}
where the coefficients $d_{m}$ are of the order $\epsilon$.

The normalization condition of eigenvectors leads therefore to
\begin{equation}
	\braket{\nu_{n} | \nu_{n}} = 1 \quad \Leftrightarrow \quad | c_{n}|^2 + \sum_{m \neq n} |
	d_{m}|^2 = 1
\end{equation}
which results in $ c_{n}  = 1- \bigO(\epsilon^2)$.

By inserting these results into the eigenvalue equation we obtain
\begin{equation}
	\left( \matM - \lambda_{n}  \right) \ket{\nu_{n}^{0}} = (\matM^{(0)}  + \epsilon
	\matM^{(1)}  - \lambda_{n}) \left[ \ket{\nu_{n}^{0} }  + \sum_{m \neq n} d_{m}
	\ket{\nu_{m}^{0}}      \right] + \bigO(\epsilon^2)
	\label{eq:evalEQ}
\end{equation}
Projection of this equation onto an eigenvector $\ket{\nu_{n'}^{0}}$ with $n'
\neq n$ leads to
\begin{equation}
	\fl \braket{\nu_{n'}^{0} | \matM - \lambda_{n} | \nu_{n}^{0}} = \epsilon
	\braket{\nu_{n'}^{0} | M_{1} | \nu_{n}^{0}}
	+ \sum_{m \neq n } d_{m} \left[
		\lambda_{m}^{0} \delta_{m,n'} + \epsilon \braket{\nu_{n'}^{0} |
			\matM^{(1)}  |
	\nu_{m}^{0}} \!-\! \lambda_{n} \delta_{m,n'}    \right] \overset{!}{=}
	0
\end{equation}
The leading order of this equation in $\epsilon$ reads
\begin{equation}
	\epsilon \braket{\nu_{n'}^{0} | \matM^{(1)}  | \nu_{n}^{0}}  +d_{n'} \left(
		\lambda_{n'}^{0} - \lambda_{n}    \right) = 0
\end{equation}
from which we readily obtain the leading order of the coefficients as
\begin{eqnarray}
	d_{n'} &= \epsilon \frac{\braket{\nu_{n'}^{0} | \matM^{(1)}  | \nu_{n}^{0}}}{\lambda_{n} -
\lambda_{n'}^{0}   } +  \bigO(\epsilon^2) \nonumber\\
		   &= \epsilon \frac{\braket{\nu_{n'}^{0} | \matM^{(1)}  | \nu_{n}^{0}}
	}{\lambda_{n}^{0} - \lambda_{n'}^{0}    }  + \bigO(\epsilon^{2})\,.
	\label{eq:evecCorr}
\end{eqnarray}
Here we could replace $\lambda_{n}$ with $\lambda_{n}^{0}$ in the second step
since the difference of the two is of order $\epsilon$ and therefore negligible
in the leading order.

By projection of eq.~\eqref{eq:evalEQ} onto $\ket{\nu_{n}^{0}}$ and by using
the results form~\eqref{eq:evecCorr}, it is now possible to solve for the
corrections to the eigenvalues up to order $\epsilon^{2} $
\begin{equation}
	\fl 0
% =\braket{\nu_{n}^{0} | M- \lambda_{n} | \nu_{n}}
	 =  \bra{\nu_{n}^{0}}  \left( \matM^{(0)}  + \epsilon \matM^{(1)}  - \lambda_{n}    \right)
	\left( \ket{\nu_{n}^{0}} + \epsilon \sum_{m \neq n}
		\frac{\braket{\nu_{m}^{0}| \matM^{(1)}  | \nu_{n}^{0} }    }{\lambda_{n}^{0} -
		\lambda_{m}^{0}  } \ket{\nu_{m}^{0}} + \bigO(\epsilon^2)
	\right)\\
\end{equation}
Solving for $\lambda_{n}$ finally yields
\begin{equation}
	\fl \lambda_{n} = \lambda_{n}^{0} + \epsilon \braket{\nu_{n}^{0}| \matM^{(1)}  |
		\nu_{n}^{0}} + \epsilon^2 \sum_{m \neq n} \frac{\braket{\nu_{n}^{0}|
			\matM^{(1)}  |
		\nu_{m}^{0}}\braket{\nu_{m}^{0} | \matM^{(1)}  | \nu_{n}^{0}}
}{\lambda_{n}^{0} - \lambda_{m}^{0}    }   + \bigO(\epsilon^3)
\end{equation}
which resembles closely the result used in quantum mechanics; the only
difference being that in this more general case the matrix elements and
eigenvalues can be complex.

\section*{References}%
\label{sec:references}

\bibliographystyle{iopart-num}
\bibliography{spectrum_bound}

\providecommand{\newblock}{}
\begin{thebibliography}{10}
\expandafter\ifx\csname url\endcsname\relax
  \def\url#1{{\tt #1}}\fi
\expandafter\ifx\csname urlprefix\endcsname\relax\def\urlprefix{URL }\fi
\providecommand{\eprint}[2][]{\url{#2}}

\bibitem{agaev_spectra_2005}
Agaev R and Chebotarev P 2005 {\em Linear Algebra and its Applications\/} {\bf
  399} 157--168

\bibitem{meyer_matrix_2000}
Meyer C~D 2000 {\em Matrix analysis and applied linear algebra\/}
  ({Philadelphia}: {Society for Industrial and Applied Mathematics}) ISBN
  978-0-89871-454-8 oCLC: 702199335

\bibitem{maroulas_perron-frobenius_2002}
Maroulas J, Psarrakos P~J and Tsatsomeros M~J 2002 {\em Linear Algebra and its
  Applications\/} {\bf 348} 49--62

\bibitem{pillai_perron-frobenius_2005}
Pillai S~U, Suel T and {Seunghun Cha} 2005 {\em IEEE Signal Processing
  Magazine\/} {\bf 22} 62--75

\bibitem{seifert_stochastic_2019}
Seifert U 2019 {\em Annu. Rev. Condens. Matter Phys.\/} {\bf 10} 171--192

\bibitem{barato_thermodynamic_2015}
Barato A~C and Seifert U 2015 {\em Phys. Rev. Lett.\/} {\bf 114} 158101

\bibitem{gingrich_dissipation_2016}
Gingrich T~R, Horowitz J~M, Perunov N and England J~L 2016 {\em Phys. Rev.
  Lett.\/} {\bf 116} 120601

\bibitem{horowitz_proof_2017a}
Horowitz J~M and Gingrich T~R 2017 {\em Phys. Rev. E\/} {\bf 96} 020103

\bibitem{proesmans_discrete-time_2017a}
Proesmans K and {Van den Broeck} C 2017 {\em EPL\/} {\bf 119} 20001

\bibitem{fei_design_2018}
Fei C, Cao Y, Ouyang Q and Tu Y 2018 {\em Nature Communications\/} {\bf 9} 1434

\bibitem{nguyen_phase_2018}
Nguyen B, Seifert U and Barato A~C 2018 {\em J. Chem. Phys.\/} {\bf 149} 045101

\bibitem{owen_number_2019}
Owen J~A, Kolchinsky A and Wolpert D~H 2019 {\em New J. Phys.\/} {\bf 21}
  013022

\bibitem{barato_coherence_2017}
Barato A~C and Seifert U 2017 {\em Phys. Rev. E\/} {\bf 95} 062409

\bibitem{dmitriev_characteristic_1945}
Dmitriev N and Dynkin E 1945 {\em C.R. (Dokl.) Acad.Sci. URSS (N.S.)\/} {\bf
  49} 159--162

\bibitem{dmitriev_characteristic_1946}
Dmitriev N and Dynkin E 1946 {\em Izvestia Akad. Nauk SSSR Ser. Mat.\/} {\bf
  10} 167--184

\bibitem{swift_location_1972}
Swift J 1972 {\em Location of {{Characteristic Roots}} of {{Stochastic
  Matrices}}\/} Master {{Thesis}} McGill University {Montreal}

\bibitem{pietzonka_universal_2016}
Pietzonka P, Barato A~C and Seifert U 2016 {\em Phys. Rev. E\/} {\bf 93} 052145

\bibitem{kraft_software_1988a}
Kraft D 1988 A software package for sequential quadratic programming Tech.
  {{Rep}}. {{DFVLR}}-{{FB}} 88-28 {DLR German Aerospace Center \textendash{}
  Institute for Flight Mechanics,} {K{\"o}n, Germany} oCLC: 27846848

\bibitem{peixoto_graph-tool_2017}
Peixoto T~P 2017 The graph-tool python library figshare

\bibitem{CSTN-013}
Hawick K~A and James H~A 2008 Enumerating {{Circuits}} and {{Loops}} in
  {{Graphs}} with {{Self}}-{{Arcs}} and {{Multiple}}-{{Arcs}} {\em Proc. 2008
  {{Int}}. {{Conf}}. on {{Foundations}} of {{Computer Science}} ({{FCS}}'08)\/}
  ({Las Vegas, USA}: {CSREA / Massey University}) pp 14--20

\end{thebibliography}

\end{document}